%% file: paper.tex
\documentclass[12pt]{article}
\usepackage{epsfig}
\usepackage{amsmath}
\usepackage{hhline}
\usepackage{amssymb}
\usepackage{times}
\usepackage{cite}

\newlength{\dinwidth}
\newlength{\dinmargin}
\begin{document}  
% The rest
\newcommand{\pom}{{I\!\!P}}
\newcommand{\reg}{{I\!\!R}}
\newcommand{\slowpi}{\pi_{\mathit{slow}}}
\newcommand{\fiidiii}{F_2^{D(3)}}
\newcommand{\fiidiiiarg}{\fiidiii\,(\beta,\,Q^2,\,x)}
\newcommand{\n}{1.19\pm 0.06 (stat.) \pm0.07 (syst.)}
\newcommand{\nz}{1.30\pm 0.08 (stat.)^{+0.08}_{-0.14} (syst.)}
\newcommand{\fiidiiiful}{F_2^{D(4)}\,(\beta,\,Q^2,\,x,\,t)}
\newcommand{\fiipom}{\tilde F_2^D}
\newcommand{\ALPHA}{1.10\pm0.03 (stat.) \pm0.04 (syst.)}
\newcommand{\ALPHAZ}{1.15\pm0.04 (stat.)^{+0.04}_{-0.07} (syst.)}
\newcommand{\fiipomarg}{\fiipom\,(\beta,\,Q^2)}
\newcommand{\pomflux}{f_{\pom / p}}
\newcommand{\nxpom}{1.19\pm 0.06 (stat.) \pm0.07 (syst.)}
\newcommand {\gapprox}
   {\raisebox{-0.7ex}{$\stackrel {\textstyle>}{\sim}$}}
\newcommand {\lapprox}
   {\raisebox{-0.7ex}{$\stackrel {\textstyle<}{\sim}$}}
\def\gsim{\,\lower.25ex\hbox{$\scriptstyle\sim$}\kern-1.30ex%
\raise 0.55ex\hbox{$\scriptstyle >$}\,}
\def\lsim{\,\lower.25ex\hbox{$\scriptstyle\sim$}\kern-1.30ex%
\raise 0.55ex\hbox{$\scriptstyle <$}\,}
\newcommand{\pomfluxarg}{f_{\pom / p}\,(x_\pom)}
\newcommand{\dsf}{\mbox{$F_2^{D(3)}$}}
\newcommand{\dsfva}{\mbox{$F_2^{D(3)}(\beta,Q^2,x_{I\!\!P})$}}
\newcommand{\dsfvb}{\mbox{$F_2^{D(3)}(\beta,Q^2,x)$}}
\newcommand{\dsfpom}{$F_2^{I\!\!P}$}
\newcommand{\gap}{\stackrel{>}{\sim}}
\newcommand{\lap}{\stackrel{<}{\sim}}
\newcommand{\fem}{$F_2^{em}$}
\newcommand{\tsnmp}{$\tilde{\sigma}_{NC}(e^{\mp})$}
\newcommand{\tsnm}{$\tilde{\sigma}_{NC}(e^-)$}
\newcommand{\tsnp}{$\tilde{\sigma}_{NC}(e^+)$}
\newcommand{\st}{$\star$}
\newcommand{\sst}{$\star \star$}
\newcommand{\ssst}{$\star \star \star$}
\newcommand{\sssst}{$\star \star \star \star$}
\newcommand{\tw}{\theta_W}
\newcommand{\sw}{\sin{\theta_W}}
\newcommand{\cw}{\cos{\theta_W}}
\newcommand{\sww}{\sin^2{\theta_W}}
\newcommand{\cww}{\cos^2{\theta_W}}
\newcommand{\trm}{m_{\perp}}
\newcommand{\trp}{p_{\perp}}
\newcommand{\trmm}{m_{\perp}^2}
\newcommand{\trpp}{p_{\perp}^2}
\newcommand{\alp}{\alpha_s}

\newcommand{\alps}{\alpha_s}
\newcommand{\sqrts}{$\sqrt{s}$}
\newcommand{\LO}{$O(\alpha_s^0)$}
\newcommand{\Oa}{$O(\alpha_s)$}
\newcommand{\Oaa}{$O(\alpha_s^2)$}
\newcommand{\PT}{p_{\perp}}
\newcommand{\JPSI}{J/\psi}
\newcommand{\sh}{\hat{s}}
\newcommand{\uh}{\hat{u}}
\newcommand{\MP}{m_{J/\psi}}
\newcommand{\PO}{I\!\!P}
\newcommand{\xbj}{x}
\newcommand{\xpom}{x_{\PO}}
\newcommand{\ttbs}{\char'134}
\newcommand{\xpomlo}{3\times10^{-4}}  
\newcommand{\xpomup}{0.05}  
\newcommand{\dgr}{^\circ}
\newcommand{\pbarnt}{\,\mbox{{\rm pb$^{-1}$}}}
\newcommand{\gev}{\,\mbox{GeV}}
\newcommand{\WBoson}{\mbox{$W$}}
\newcommand{\fbarn}{\,\mbox{{\rm fb}}}
\newcommand{\fbarnt}{\,\mbox{{\rm fb$^{-1}$}}}
%
% Some useful tex commands
%
\newcommand{\qsq}{\ensuremath{Q^2} }
\newcommand{\gevsq}{\ensuremath{\mathrm{GeV}^2} }
\newcommand{\et}{\ensuremath{E_t^*} }
\newcommand{\rap}{\ensuremath{\eta^*} }
\newcommand{\gp}{\ensuremath{\gamma^*}p }
\newcommand{\dsiget}{\ensuremath{{\rm d}\sigma_{ep}/{\rm d}E_T^*} }
\newcommand{\dsigrap}{\ensuremath{{\rm d}\sigma_{ep}/{\rm d}\eta^*} }
\newcommand{\etgap}{E_T^{gap}}
\newcommand{\etcut}{E_T^{cut}}
\newcommand{\xgjet}{x_{\gamma}^{jets}}
\newcommand{\xpjet}{x_{p}^{jets}}
\def\pythia{{\sc Pythia}}
\def\herwig{{\sc Herwig}}
\def\jimmy{{\sc Jimmy}}
\def\jetset{{\sc Jetset}}
\newcommand{\deta}{\Delta \eta}
% Journal macro
\def\Journal#1#2#3#4{{#1} {\bf #2} (#3) #4}
\def\NCA{\em Nuovo Cimento}
\def\NIM{\em Nucl. Instrum. Methods}
\def\NIMA{{\em Nucl. Instrum. Methods} {\bf A}}
\def\NPB{{\em Nucl. Phys.}   {\bf B}}
\def\PLB{{\em Phys. Lett.}   {\bf B}}
\def\PRL{\em Phys. Rev. Lett.}
\def\PRD{{\em Phys. Rev.}    {\bf D}}
\def\ZPC{{\em Z. Phys.}      {\bf C}}
\def\EJC{{\em Eur. Phys. J.} {\bf C}}
\def\CPC{\em Comp. Phys. Commun.}

\begin{titlepage}

\noindent
DESY 02-023  \hfill  ISSN 0418-9833 \\
March 2002

\vspace*{3cm}

%Date: Feb 26 2002              \\
%Version: 3.0           \\
%Editors: B. Cox and A. Wyatt           \\
%Referees: S. Maxfield and J. Turnau          \\   
%\vspace{2cm}

\begin{center}
\begin{Large}

{\bf Energy Flow and Rapidity Gaps Between Jets in} \\
{\bf Photoproduction at HERA}

\vspace*{1cm}

{\Large H1 Collaboration}

\end{Large}
\end{center}

\vspace*{3cm}

\begin{abstract}
Dijet events in photon-proton collisions in which there is a large pseudorapidity separation $\Delta \eta > 2.5$ between the two highest $E_{\hspace{2pt}T}$ jets  are studied with the H1 detector at HERA.  
The inclusive dijet cross sections are measured as functions of the longitudinal momentum fractions of the proton and photon which participate in the production of the jets, $\xpjet$ and $\xgjet$ respectively, $\Delta \eta$, the pseudorapidity separation between the two highest $E_T$ jets, and $\etgap$, the total summed transverse energy between the jets. Rapidity gap events are defined as events in which $\etgap$ is less than $\etcut$, for $E_T^{cut}$ varied between $0.5$ and $2.0$ GeV. The fraction of dijet events with a rapidity gap is measured differentially in $\Delta \eta$, $\xpjet$ and $\xgjet$. An excess of events with rapidity gaps at low values of  $\etcut$ is observed above the expectation from standard photoproduction processes. This excess can be explained by the exchange of a strongly interacting colour singlet object between the jets.
\end{abstract}

\vspace{1.5cm}

\begin{center}
To be submitted to Eur. Phys. J. 
\end{center}

\end{titlepage}

\begin{flushleft}
  \input{h1auts}
\end{flushleft}

\newpage

\section{Introduction}
\noindent
\label{intro}
\begin{figure}
\begin{center}
\epsfig{file=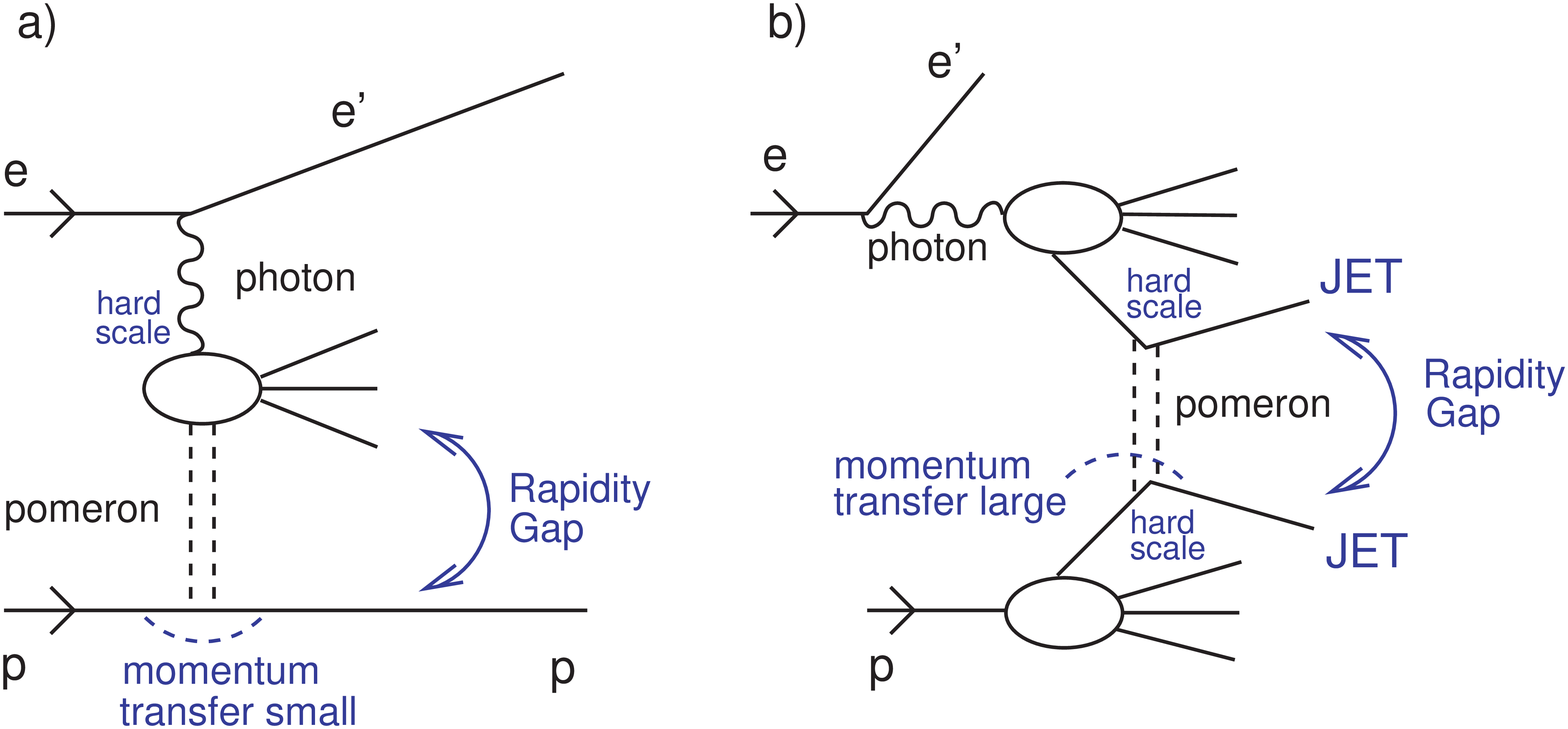,width=.9\textwidth}
\end{center}
\caption{The diffractive deep inelastic scattering process (a) compared with the rapidity gaps between jets process (b) at HERA.}         
\label{feynman}
\end{figure}
Events with large rapidity gaps in the hadronic final state coupled with the presence of a hard scale have long been recognized as an ideal place to study the interplay between long and short distance physics \cite{FR}. Measurements of diffractive deep inelastic scattering at HERA \cite{H1diff,ZEUSdiff}, for example, have led to a phenomenology which mixes the languages of perturbative quantum chromodynamics (QCD) and Regge theory. In this case the proton remains intact or dissociates into a low mass system: the momentum transfer across the gap is small. In other words there is no hard scale at the proton vertex 
(see figure \ref{feynman}(a)). A special class of events in which there is a large momentum transfer across the gap has been observed in proton - antiproton collisions at the Tevatron \cite{DZero1,DZero2,DZero3,CDF1,CDF2,CDF3} and is the subject of this study. Such events are characterised by two high $E_T$ jets in the final state separated by a large rapidity interval and little or no energy flow between the jets 
(see figure \ref{feynman}(b)). At such large momentum transfers the Regge-inspired phenomenology of diffractive deep inelastic scattering predicts cross sections that are orders of magnitude smaller than those measured. In a $p \bar p$ collision, the process may be visualised as the hard scattering of one parton from each proton via the exchange of a strongly interacting colour singlet object, the ``perturbative pomeron''. At HERA the ZEUS Collaboration observed this process in photoproduction events, in which a quasi real photon emitted from the incoming positron interacts with the proton \cite{zeus}. The photon can interact either as a point-like object, which in leading order QCD is termed the direct process, or it can fluctuate into a hadronic state from which a parton can interact with a parton in the proton, the resolved process. In resolved events, parton - parton hard scattering via the exchange of a colour singlet object should be present, and it is one of the aims of this paper to search for these events.

It might be expected that the rate for such a process should be calculable using only perturbative methods, since the presence of a large momentum transfer ensures that the gap production mechanism is dominated by short distance physics \cite{FS}: the perturbative pomeron couples to individual partons. The presence of a rapidity gap together with the large rapidity separation of the jets and large momentum transfer means that one is in a perturbatively calculable Regge limit of QCD, i.e. $\hat s \gg -\hat t \gg \Lambda_{QCD}^2$, where $\hat t$ is the momentum transfer across the gap and $\hat s$ is the parton - parton centre of mass energy. Calculations using the leading logarithmic approximation (LLA) of BFKL \cite{BFKL1,BFKL2,BFKL3} have been performed \cite{MT,enberg}. 
They have been found to describe an observed excess of dijet events with low levels of hadronic activity between the jets, over the expectation from standard QCD processes, at the Tevatron \cite{CFL,enberg}. 
The situation is complicated, however, by the possibility that interactions between spectator partons in the colliding hadrons can destroy the gap \cite{Bj,fsel,GLM,Kaidalov:2001iz}. 
This introduces a potentially large non-perturbative component into such calculations, making absolute predictions of both rates and differential distributions problematic. 
It has been noted by several authors \cite{OS,OShera,CFL} that this problem can be avoided, and perturbative predictions are in principal possible, if a rapidity gap is defined in terms of the energy $\etgap \gg \Lambda_{QCD}$ in the pseudorapidity region between the jets. 
It is the aim of this paper to search for dijet events with rapidity gaps in a way which will allow such theoretical calculations to be compared to the experimental data in the future. 
An alternative approach to BFKL calculations is offered by the Colour 
Evaporation Model, in which the gap formation mechanism itself is considered 
to be purely non-perturbative. This model has also been shown to describe 
the Tevatron and HERA data \cite{Eboli:1999dd}.

After a brief description of the H1 detector in section \ref{detector}, the 
definition of a rapidity gap event and the selection of the event sample are 
discussed in section \ref{evselect}. The Monte Carlo models used to correct 
the data and to compare to the BFKL predictions for the rapidity gap 
production rates are described in section \ref{mc}. In section \ref{results} 
the data are presented, and the results are discussed in 
section \ref{discuss}. 

\section{The H1 Detector}
\label{detector}
A detailed description of the H1 apparatus can be found 
elsewhere~\cite{h1nim}. 
The following briefly describes the detector components relevant to this
analysis.

A liquid argon (LAr)
calorimeter covers the range in polar angle  
$4^{\circ} < \theta < 153^{\circ}$ $(3.35 > \eta > -1.43)$ with full azimuthal
coverage\footnote{$\theta$ is measured relative to the outgoing proton beam direction which defines the positive $z$ axis and the forward direction. Pseudo-rapidity is defined as $\eta = -{ \rm ln (tan} \theta / 2)$.}. The LAr calorimeter consists of an electromagnetic section with lead absorbers
 and a hadronic section with steel absorbers, of 
combined depth between 4.5 and 8 interaction lengths. Both sections
are highly segmented in the
transverse and longitudinal directions with about 44000 cells in total.
The absolute hadronic energy scale is known to $4 \%$
for this analysis. The polar region 
$153^{\circ} < \theta < 177.8^{\circ}$ $(-1.43 > \eta > -3.95)$ is covered by the SPACAL \cite{spac}, a 
lead-scintillating fibre calorimeter with both electromagnetic and hadronic 
sections, with a combined depth of 2 interaction lengths. The hadronic energy scale uncertainty is presently 
$7 \%$.  
Tracking information is provided by the two concentric drift chambers of the central tracker, covering the pseudo-rapidity range $-1.5 < \eta < 1.5$, which lie inside a $1.15$ T solenoidal field. 

The luminosity 
is measured from the reaction $ep \rightarrow ep\gamma $ with two TlCl/TlBr 
crystal calorimeters \cite{h1nim}, the electron and photon taggers, installed in the HERA tunnel. The electron 
tagger is 
located at $z = -33 \ {\rm m}$ from the interaction 
point in the direction of the 
outgoing lepton beam and the photon tagger at $z = -103 \ {\rm m}$.

\section{Event Selection and Kinematic Reconstruction}
\label{evselect}
The data for this analysis were collected with the H1 detector during the 
1996 running period, when HERA collided 27.6 {\rm GeV} positrons with 820 {\rm GeV} 
protons.  An integrated luminosity of 6.6 ${\rm pb}^{-1}$ is used.
Photoproduction events were selected by detecting the scattered positron 
in the electron tagger of the luminosity system. 
This 
restricts the virtuality of the photon to $Q^{2} < 0.01 $ GeV$^{2}$. The
photon-proton centre of mass energy was restricted to the range
$165 < W < 233 \ {\rm GeV}$ to avoid regions of low electron 
tagger acceptance. 

Two triggers were used in the analysis, one based on tracking requirements in the central tracker, the other on energy in the electromagnetic section of the SPACAL calorimeter. The combined efficiency of the two triggers, was greater than $60 \%$ for all measured data points. Both triggers also required an electron to be detected in the electron tagger. 
 
Hadronic final state objects were defined using a combination of tracking and calorimetric information. An algorithm was used which avoids the double counting of tracks and calorimeter clusters \cite{fscomb}.
The inclusive $k_\perp$ clustering algorithm was applied to the hadronic final state objects. The algorithm was run in the `covariant $p_T$ scheme' with the R parameter set to 1 \cite{Ellis:tq,kt}. In this inclusive mode, the $k_\perp$ algorithm merges every final state object in the event, excluding the scattered electron, uniquely into a list of massless jets, ordered in $E_T$. Events were kept if the two highest $E_T$ jets in the event satisfied the following criteria:
\begin{eqnarray}
E_T^{jet,1} > 6.0 ~\rm{GeV} \\
E_T^{jet,2} > 5.0 ~\rm{GeV} \\
\eta^{jet,1},\eta^{jet,2} < 2.65 \\
2.5 < \Delta \eta \equiv |\eta^{jet,1}-\eta^{jet,2}| < 4.0
\end{eqnarray}
The first two cuts are asymmetric to avoid regions in which NLO
QCD predictions become unstable \cite{frixione}, in order 
to facilitate the comparison
of the data with such calculations when they become available.
%The first two cuts are asymmetric in order to facilitate comparison with NLO QCD calculations, when such calculations become available. 
The third cut ensures that the most forward jet is well contained within the LAr calorimeter. The fourth cut ensures a large jet-jet separation. No backward $\eta$ cut on the jets was necessary, as the other kinematic cuts force the backward jet to be within the acceptance of the H1 detector.

The total transverse energy between the two highest $E_T$ jets, $\etgap$, is defined as the sum of the energy of all jets whose pseudorapidities, $\eta^{jet,i}$, lie between their axes, i.e.
\begin{equation}
\etgap = \sum_{i>2} E_T^{\mit{jet,i}} \hspace{.2cm}, \hspace{.2cm}\eta^{jet}_{\mit{forward}} > \eta^{jet,i} > \eta^{jet}_{\mit{backward}}.
\end{equation} 
 
In addition to the kinematic variables defined above, two variables $\xgjet$ and $\xpjet$ are defined as the fractional longitudinal momentum of the photon and proton participating in the production of the two highest $E_T$ jets  
\begin{eqnarray}
\xgjet = \frac{\sum_{i=1,2} \left(E^{jet,i}-p_z^{jet,i}\right)}{\sum_{obj}(E^{obj}-p_z^{obj})} \\
\xpjet =  \frac{\sum_{i=1,2} \left(E^{jet,i}+p_z^{jet,i}\right)}{2E_p}
\end{eqnarray}
where $E_p$ is the energy of the incoming proton and the sum in the denominator of the $\xgjet$ calculation runs over all hadronic final state objects in the event, excluding the scattered electron.

\section{Monte Carlo simulations and data corrections}
\label{mc}

The \pythia~5.7 \cite{PYTHIA} and \herwig~6.1 \cite{HERWIG} Monte Carlo event generators were used to correct the data for detector acceptance and bin migration effects, and for model comparisons. Both generators simulate the direct and resolved production of dijets by quasi real photons. The hard scattering matrix elements are calculated to leading order, regulated by a cut-off, $P_T^{min}$. In addition to the primary hard scatter, both generators contain models to simulate the effects of multiple parton-parton interactions in a resolved photoproduction event. In \pythia, the probability to have several parton-parton collisions in a single event is modeled using the parton densities and the usual leading order matrix elements \cite{PYTHIA}. As for the primary hard scatter, the matrix elements are divergent and must be regulated by a cut-off, $P_T^{mp}$, which is the main free parameter in the model. This parameter was tuned to give the best description of the H1 data, after all kinematic cuts, used in this analysis\footnote{The simplest multiple interaction model in \pythia~was used, corresponding to setting switch \texttt{MSTP(82)=1}.}. In the case of \herwig, multiple interactions were simulated by the \jimmy~package \cite{jimmy}.     
  The main free parameter in the \herwig~$+$ \jimmy~model is  $P_T^{mp}$, as for \pythia, which in the current version of \jimmy~must be set equal to $P_T^{min}$, the cut off for the matrix elements for the primary hard scatter. Again, this parameter was tuned to give the best description of the H1 data, after all kinematic cuts, used in this analysis.  
\begin{table}
\begin{center}
\begin{tabular}{|l|l|l|l|l|l|l|} \hline
Monte Carlo & $\alpha_s$ & Proton PDF & $\gamma$ PDF & $P_T^{min}$ (GeV) & $P_T^{mp}$ (GeV)\\ \hline
PYTHIA 5.7 & 1-loop & GRV-LO \cite{GRV} & GRV-LO \cite{Gluck:1991jc} & 2.2 & 1.5 \\ \hline
HERWIG 6.1 + JIMMY & 2-loop & GRV-LO & GRV-LO & 1.8 & 1.8  \\ \hline
\end{tabular}
\end{center}
\caption{Monte Carlo parameters.}
\label{mctable}
\end{table} 
The settings of the above parameters and parton densities used in each of the Monte Carlo simulations are shown in table \ref{mctable}. 

For both \herwig~and \pythia~the direct and resolved photoproduction processes were generated separately and added together according to their generated cross sections. The overall normalisation of the \pythia~sample was scaled by a factor of $0.7$ in order to fit the measured inclusive dijet photoproduction cross sections, shown in 
figure \ref{xs}. 
Similarly, the \herwig~sample was scaled by a factor of $1.2$. 

\herwig~incorporates the BFKL LLA colour singlet exchange cross section for the elastic scattering of two partons as computed by Mueller and Tang \cite{MT}. In the limit $\deta \gg 1$ the cross section for quark - quark scattering may be approximated as\footnote{We used a version of \herwig~in which the BFKL LLA colour singlet exchange cross section is valid for all $\deta$ \cite{CFL}.}   
\begin{equation}
\frac{d \sigma(q q \to q q)}{d \hat t} \approx (C_F \alps)^4 \frac{2\pi^3}{\hat t^2}
\frac{\exp(2 \omega_0 y)}{(7 \alps C_A \zeta(3) y)^3}
\label{mullertang}
\end{equation} 
where
\begin{equation}
y \equiv \deta = \ln \left( \frac{\hat{s}}{-\hat t} \right)
\end{equation}
and
\begin{equation}
\omega_0 = C_A (4 \ln 2/\pi) \alps.
\label{intercept}
\end{equation}
Here, $1+\omega_0$ is the perturbative pomeron intercept, $C_F = \frac{4}{3}$ is the usual colour factor for quark-quark scattering, $C_A$ is the number of colours and $\zeta$ is the Riemann $\zeta$-function. 
 The values of $\alpha_s$ in equations \ref{mullertang} and \ref{intercept} are free parameters in the LLA, and are each chosen to be 0.18. This corresponds to a choice of pomeron intercept of $1.48$. 

BFKL pomeron exchange has not yet been implemented in \pythia. Colour singlet exchange events were modeled by high-$t$ photon exchange, with a scale factor of $1200$ applied to the generated cross section. The reason for using this process is twofold; firstly, to obtain a sample of events from \pythia~in which the hadronic final state is that of a colour singlet exchange process, for the purposes of detector acceptance and bin migration corrections. Secondly, the high-$t$ photon exchange process provides a useful way of ascertaining how sensitive the data are to the underlying dynamics of colour singlet exchange, since the dynamics of photon exchange are very different to those of BFKL pomeron exchange. This will be discussed further in section \ref{results}.

The \pythia~Monte Carlo sample described above, including high-$t$ photon exchange, was passed through a full simulation of the H1 detector and used to correct the data for detector acceptance and bin migration effects. The scaled high-$t$ photon exchange sample was added to the scaled photoproduction sample to fit the measured excess of rapidity gap events. The data were also corrected using the \herwig~+ BFKL sample, and the difference assigned as a systematic error. For the purposes of detector corrections, the BFKL sample was scaled by a factor of $0.8$ and then added to the scaled photoproduction sample, again to fit the measured excess. 

The transverse energy flow around the jet axes predicted by the \herwig~and \pythia~generators is compared to the uncorrected data\footnote{The $\eta$ profile shows $\delta \eta = \eta^{\rm{object}} - \eta^{\rm{jet}}$, weighted by the object transverse energy, for objects within $1$ radian in $\phi$ of the centre of the jet. The $\phi$ profile is similarly defined for objects within $1$ unit of pseudorapidity of the jet centre.} in figure \ref{profiles}, 
for all events surviving the selection cuts described in section \ref{evselect}, 
after being passed through a complete simulation of the H1 detector.  Both \pythia~and \herwig~give a good description of the data both inside and outside the jets.

\section{Results and Model Comparisons}
\label{results}

\begin{figure}
\begin{center}
  \epsfig{file=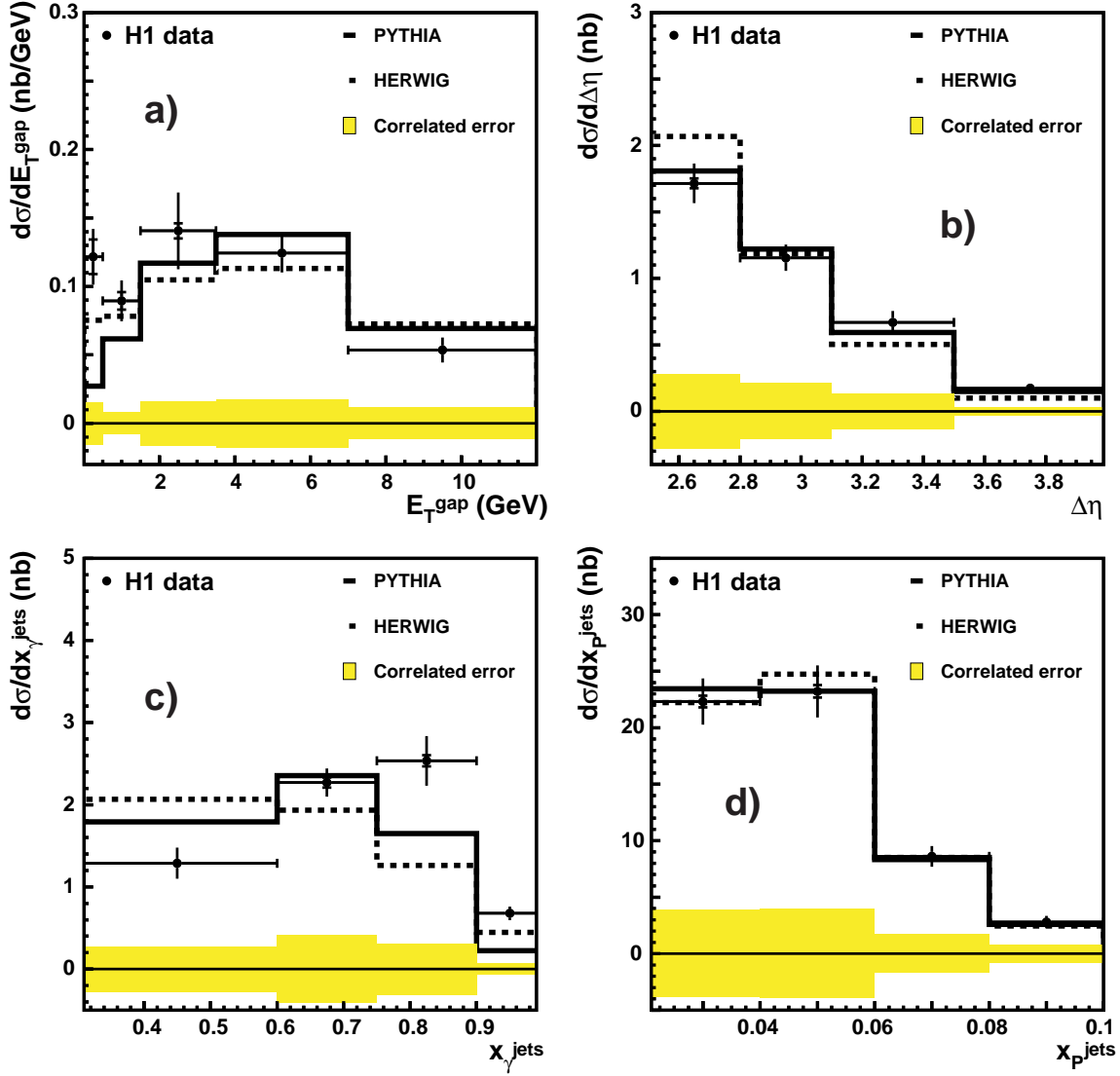,width=\textwidth}
\end{center}
\caption{In (a), the solid points show the dijet cross section differential in $\etgap$, the summed transverse energy between the two highest $E_T$ jets, in the kinematic range defined in section \ref{evselect}. The inner error bars represent the statistical error, and the outer error bars represent the statistical and non-correlated systematic errors added in quadrature. The solid band below the plot shows the correlated systematic errors, as described in the text. The dashed line shows the prediction of \herwig, scaled by a factor of $1.2$, and the solid line that of \pythia, scaled by a factor of $0.7$. In (b), (c) and (d) the dijet cross sections differential in $\Delta \eta$, $\xgjet$ and $x_p^{jets}$  are shown.\label{xs}}
\end{figure}

The $ep$ inclusive dijet cross sections, corrected for detector effects, in the kinematic range described in section \ref{evselect}, are given in 
table \ref{inctable} and shown in figure \ref{xs}. The cross sections are defined at the level of stable hadrons. The inner error bars show the statistical error and the outer error bars show the statistical and uncorrelated systematic errors added in quadrature. The dominant uncorrelated systematic errors arise from the uncertainty in the trigger efficiency ($\sim 5 \%$) and the model dependence of the Monte Carlo correction procedure ($\sim 5 \%$). The correlated systematic errors are shown as the solid band below the plots. The dominant error comes from the uncertainty in the hadronic energy scale of the LAr calorimeter. In all plots, the bin sizes are chosen to keep the effects of migrations to an acceptable level. In figure \ref{xs} the \herwig~and \pythia~Monte Carlo generator predictions are shown without additional colour singlet exchange components added. 

Figure \ref{xs}(a) shows the dijet cross section differential in $\etgap$.  For $\etgap < 0.5$ GeV there is a marked excess in the data over the prediction of both Monte Carlo generators although \herwig~predicts a larger cross section than \pythia. Figure \ref{xs}(b) shows the dijet cross section differential in $\Delta \eta$. Both \pythia~and \herwig~give  reasonable descriptions of the shape of the distribution. In figure \ref{xs}(c), the dijet cross section differential in $\xgjet$ is plotted. Both generators fail to describe the shape of the $\xgjet$ distribution. Figure \ref{xs}(d) shows the dijet cross section differential in $\xpjet$. The minimum kinematically accessible $x_p^{jets}$ is set by the requirement that $\Delta \eta > 2.5$, i.e. the parton - parton centre of mass energy must be large\footnote{The parton level variables
$x_p$ and $x_{\gamma}$ are correlated, since the rapidity separation between the outgoing partons $y = {\rm ln} (x_p x_{\gamma} W^2 / -\hat t)$.}. Both \herwig~and \pythia~describe the shape of the $x_p^{jets}$ distribution well.    

In order to look in more detail at the excess of events with low values of $\etgap$ it is helpful to look at the gap fraction, primarily because the bulk of the systematic errors cancel. The gap fraction is formed by taking the ratio of the dijet cross section, with the additional requirement that $\etgap < \etcut$, to the inclusive dijet cross section, for $\etcut =0.5$, $1.0$, $1.5$ and $2.0$ GeV.  The motivations for choosing several different values of $\etcut$ to define the gap fractions are primarily theoretical. As discussed in 
section \ref{intro}, perturbative calculations of the gap fraction are possible if $\etcut$ is chosen to be sufficiently large \cite{OS,OShera}, specifically $\sqrt{-\hat t} \gg \etcut \gg \Lambda_{QCD}$. Such calculations are able to predict the change in the gap fraction as $ \etcut$ is increased. The cross sections for events with $\etgap < \etcut$ are given differentially in $\Delta \eta$, $\xgjet$ and $\xpjet$ in tables \ref{deteatable}, \ref{xgtable} and \ref{xptable} respectively, along with the associated gap fractions. 

\begin{figure}
\begin{center}
  \epsfig{file=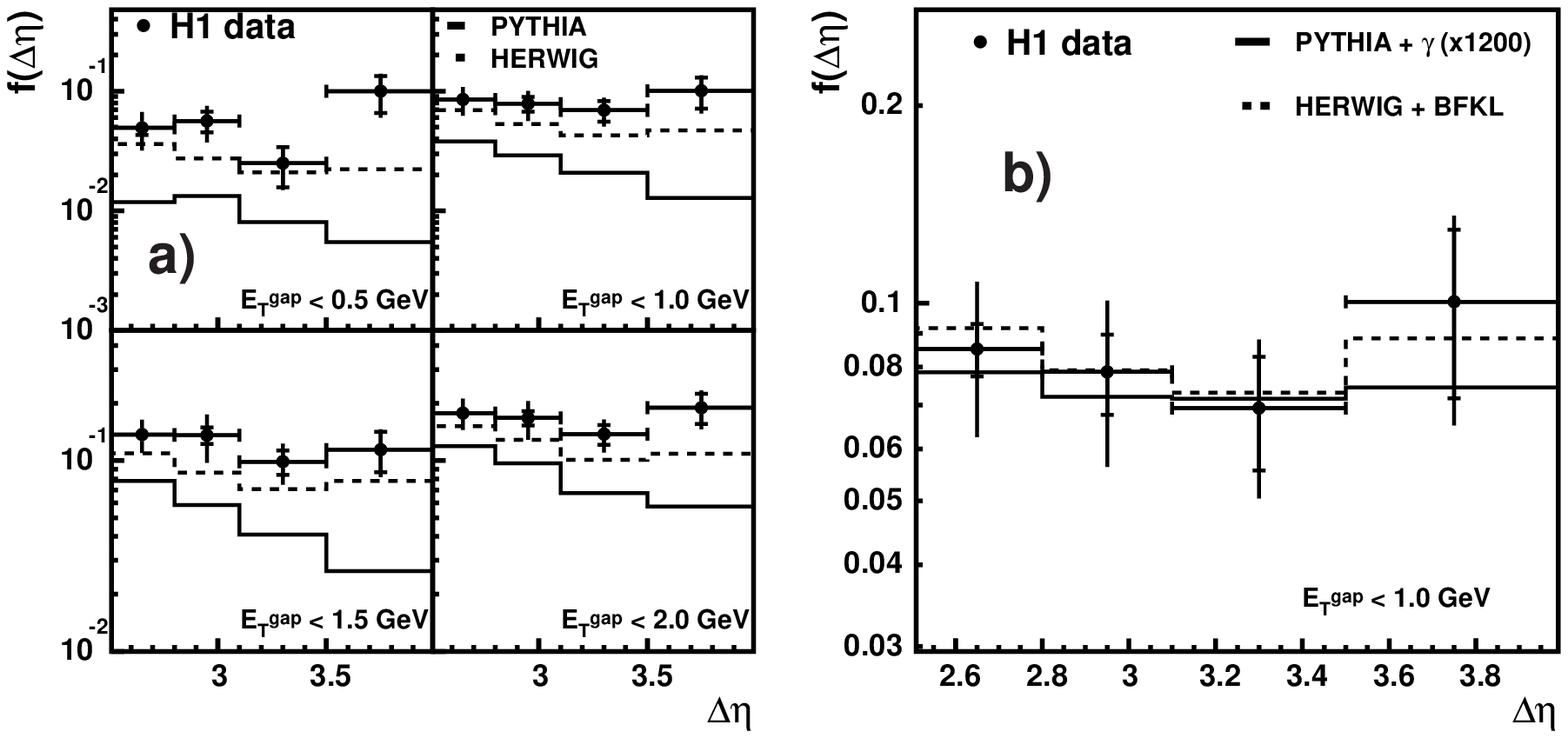,width=\textwidth}
\end{center}
\caption{The solid points show the gap fraction differential in $\Delta \eta$. The inner error bars represent the statistical error, and the outer error bars represent the statistical and all systematic errors added in quadrature. Gap events are defined for 4 values of $\etgap$, shown in the figures. In (a), the gap fractions are compared to the prediction of \herwig~(dashed line) and \pythia~(solid line). In (b), the gap fraction is shown for $\etgap < 1.0$ GeV, and compared to the \herwig~and \pythia~predictions with 2 different models of colour singlet exchange added. The dashed line shows \herwig~+ BFKL and the solid line shows \pythia~+ high-$t$ photon exchange. The photon exchange cross section is scaled by a factor of $1200$ (see text).\label{detacs}}
\end{figure}

In figure \ref{detacs}(a) the gap fraction differential in $\Delta \eta$ is shown for the 4 different choices of $\etcut$. 
Also shown are the predictions of \herwig~and \pythia. There is a clear excess over the \pythia~prediction which persists up to the largest value of $\etcut$ and increases with $\Delta \eta$. One would naively expect the gap fraction to fall exponentially with increasing $\Delta \eta$ in the absence of a colour singlet exchange component, given the assumption that multiplicity fluctuations in the hadronic final state obey Poisson statistics. This expectation is borne out by the \pythia~Monte Carlo. The data do not display this behaviour, and possibly rise in the highest $\Delta \eta$ bin. This behaviour is indicative of the presence of a colour singlet exchange  component in the data. The shape of the \herwig~distribution is much flatter than that of \pythia, and does not fall exponentially as $\Delta \eta$ increases.  \herwig~is closer to the data than \pythia,  although at large $\Delta \eta$ there is again a noticeable excess in the data for all $\etcut$. The difference between the \herwig~and \pythia~predictions is due to the different models of hadronisation in the two generators: \jetset \cite{PYTHIA} in the case of \pythia, and the cluster fragmentation model in \herwig \cite{HERWIG}. At the level of parton showering, i.e. pre-hadronisation, both generators exhibit the expected exponential fall of the gap fraction with increasing $\Delta \eta$.

In figure \ref{detacs}(b) the gap fraction for $\etgap < 1.0$ GeV is reproduced together with the predictions of \herwig~and \pythia~with colour singlet exchange models added. In this and subsequent figures, the BFKL cross section was not scaled to fit the data. The normalisation was solely determined by the choice of $\alpha_s = 0.18$ as described in section \ref{mc}. 
The \pythia~high-~$|t|$~photon exchange prediction was scaled by a factor of $1200$.
It is worth repeating here that the \pythia~model is not intended to be a candidate for the observed excess - it is not a strongly interacting process. The rationale behind including the curve is to test the sensitivity of the data to the underlying dynamics of the exchange. In particular, photons couple only to quarks, whereas a gluonic object such as the BFKL pomeron couples preferentially to gluons. As is evident from the figure, there is no significant difference in the shapes of the \herwig~and \pythia~distributions: both models are compatible with the data.

\begin{figure}
\begin{center}
  \epsfig{file=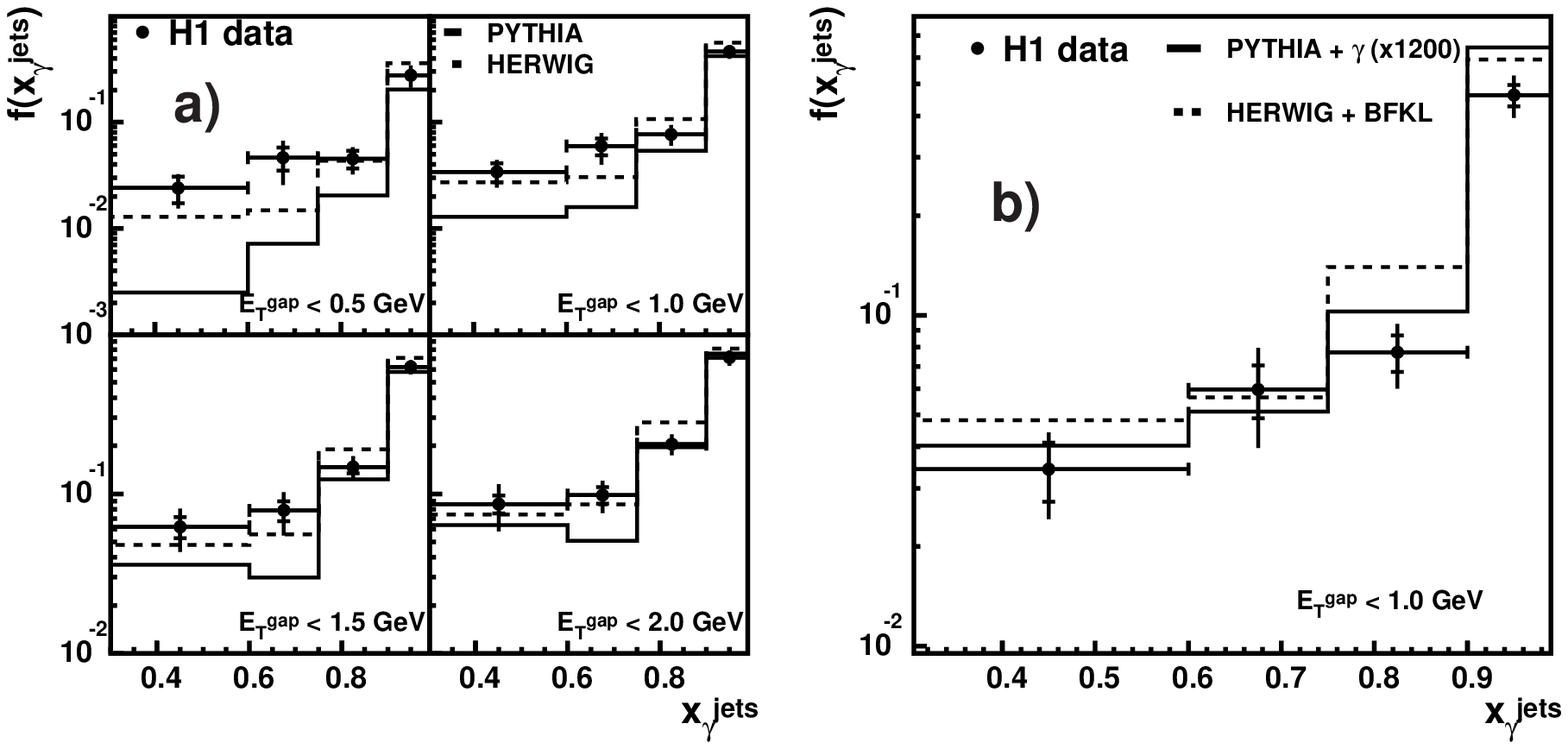,width=\textwidth}
\end{center}
\caption{The solid points show the gap fraction differential in $\xgjet$, for $\Delta \eta > 2.5$. The inner error bars represent the statistical error, and the outer error bars represent the statistical and all systematic errors added in quadrature. Gap events are defined for 4 values of $\etgap$, shown in the figures. In (a), the gap fractions are compared to the prediction of \herwig~(dashed line) and \pythia~(solid line). In (b), the gap fraction is shown for $\etgap < 1.0$ GeV, and compared to the \herwig~and \pythia~predictions with 2 different models of colour singlet exchange added. The dashed line shows \herwig~+ BFKL and the solid line shows \pythia~+ high-$t$ photon exchange. The photon exchange cross section is scaled by a factor of $1200$ (see text).\label{xgcs}}
\end{figure}

In figure \ref{xgcs}(a) the gap fraction is plotted differentially in $\xgjet$. For $\etgap < 0.5$ GeV, there is an excess visible in the data for $\xgjet < 0.75$ over the Monte Carlo predictions, although as noted above, \herwig~predicts a larger cross section at low values of $\etgap$ than \pythia. The gap fraction rises significantly at high values of $\xgjet$. This effect is reproduced by both generators, and is due to the fact that, in leading order QCD, direct photoproduction events (high $\xgjet$) have quark propagators between the outgoing partons associated with the jets, whilst the majority of resolved events (low $\xgjet$) have gluon propagators. Quark exchange diagrams lead to a lower probability of radiation in the rapidity region between the jets than gluon exchange diagrams, and hence the gap fraction increases at large $\xgjet$. 

In figure \ref{xgcs}(b) the gap fraction for $\etgap < 1.0$ GeV is reproduced together with the predictions of \herwig~and \pythia~with colour singlet exchange models added.  A better description of the gap fraction for $\xgjet < 0.75$ is achieved for both models, but at the expense of too high a gap fraction at large  $\xgjet$.    

\begin{figure}
\begin{center}
  \epsfig{file=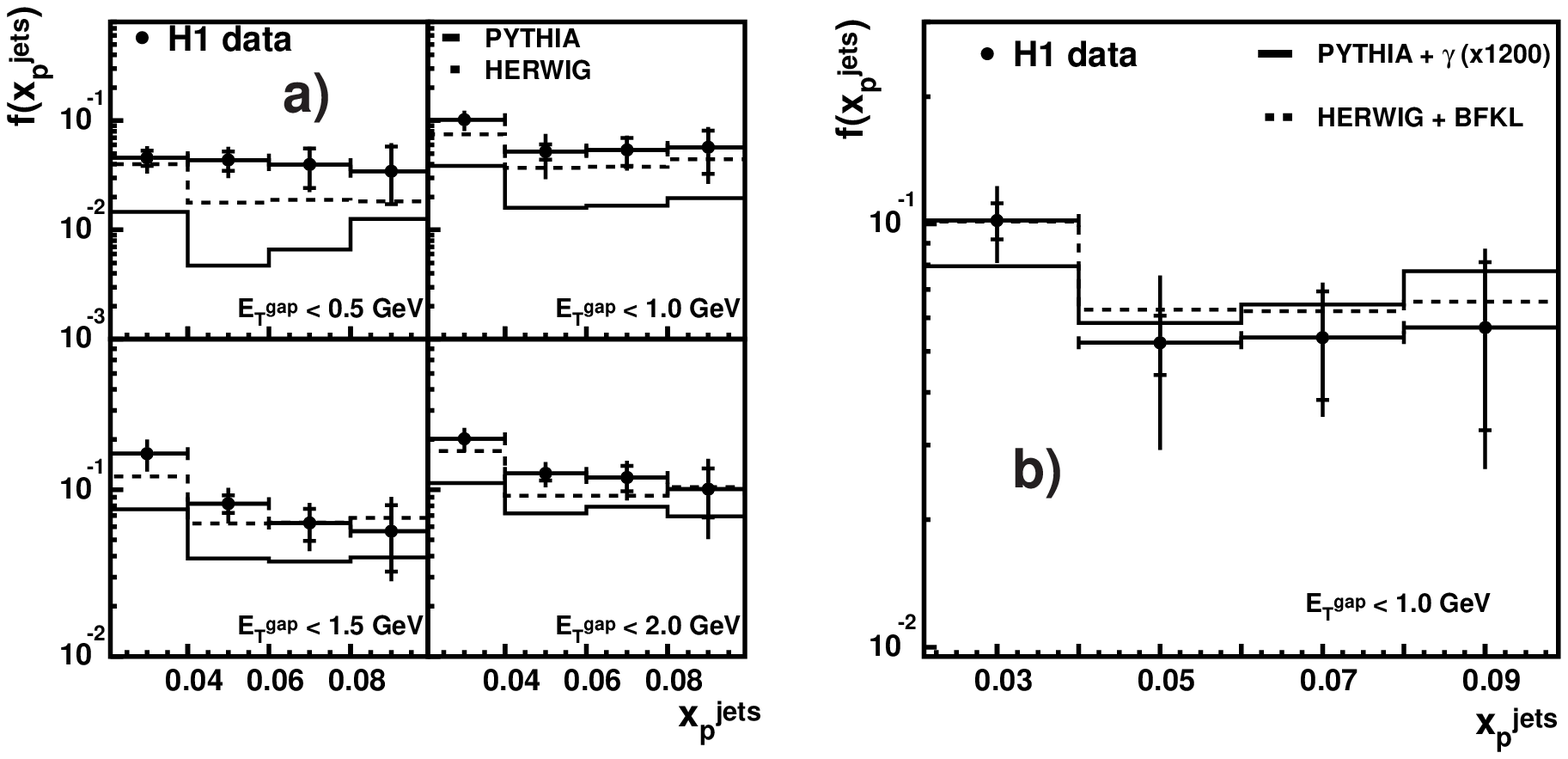,width=\textwidth}
\end{center}
\caption{The solid points show the gap fraction differential in $x_p^{jets}$, for $\Delta \eta > 2.5$. The inner error bars represent the statistical error, and the outer error bars represent the statistical and all systematic errors added in quadrature. Gap events are defined for 4 values of $\etgap$, shown in the figures. In (a), the gap fractions are compared to the prediction of \herwig~(dashed line) and \pythia~(solid line). In (b), the gap fraction is shown for $\etgap < 1.0$ GeV, and compared to the \herwig~and \pythia~predictions with 2 different models of colour singlet exchange added. The dashed line shows \herwig~+ BFKL and the solid line shows \pythia~+ high-$t$ photon exchange. The photon exchange cross section is scaled by a factor of $1200$ (see text).\label{xpcs}}
\end{figure}

In figure \ref{xpcs}(a) the gap fraction differential in $x_p^{jets}$ is shown. Again, there is an excess over the \pythia~prediction which persists up to the largest value of $\etcut$. In this case, the excess is present in all bins. The tendency for the data and Monte Carlo predictions to rise at low $x_p^{jets}$ is due to the correlation between $x_p^{jets}$ and $\xgjet$: for large $\Delta \eta$, small $x_p^{jets}$ must be compensated by large $\xgjet$. The excess over \herwig~is less pronounced. In figure \ref{xpcs}(b), the gap fraction for $\etgap < 1.0$ GeV is replotted with the predictions of \herwig~and \pythia~with colour singlet models added. Both models are able to describe the data.

\section{Discussion}
\label{discuss}

It is clear from the gap fractions presented in figures \ref{detacs}, \ref{xgcs} and \ref{xpcs} that there is a highly significant excess of events with low energy flow between the jets over that predicted by the \pythia~generator before the addition of a colour singlet exchange model. The measurements also show an excess over the predictions of \herwig, although the effect is less pronounced. The gap fraction in the range $3.5 < \Delta \eta < 4.0$ for $\etgap < 0.5$ GeV is approximately $10\%$. This provides the closest comparison with the result of the ZEUS Collaboration, who found a gap fraction of approximately $11\%$ at $\Delta \eta = 3.7$, but in a slightly different kinematic range and with a different definition of a rapidity gap \cite{zeus}. 

The large 4-momentum transfer across the rapidity gap forced by the selection of high transverse energy jets means that standard Regge inspired phenomenology cannot explain this excess. One particular solution to this discrepancy has been investigated here, namely the addition of a distinct hard colour singlet exchange component to the simulations. The two different models of the underlying dynamics of the colour singlet component considered here could not be distinguished by the data. Whilst it is interesting that the LLA BFKL cross section is of the right order of magnitude to fit the data for a choice of $\alpha_s = 0.18$, there are significant uncertainties which make strong conclusions difficult to draw.  Equation \ref{mullertang} has been derived using the leading logarithmic approximation of BFKL, and the higher order corrections at non-zero $t$ are unknown at present. Even within the LLA, there are large uncertainties due to the choice of scales, the treatment of the running coupling (which has been treated as a fixed parameter in this analysis) and the contribution of higher conformal spin \cite{enberg}. A further complication is introduced by the issue of rapidity gap survival; colour singlet exchange events will not necessarily all lead to rapidity gaps. This has not been taken into account in the BFKL predictions shown in this paper.

\begin{figure}
\begin{center}
  \epsfig{file=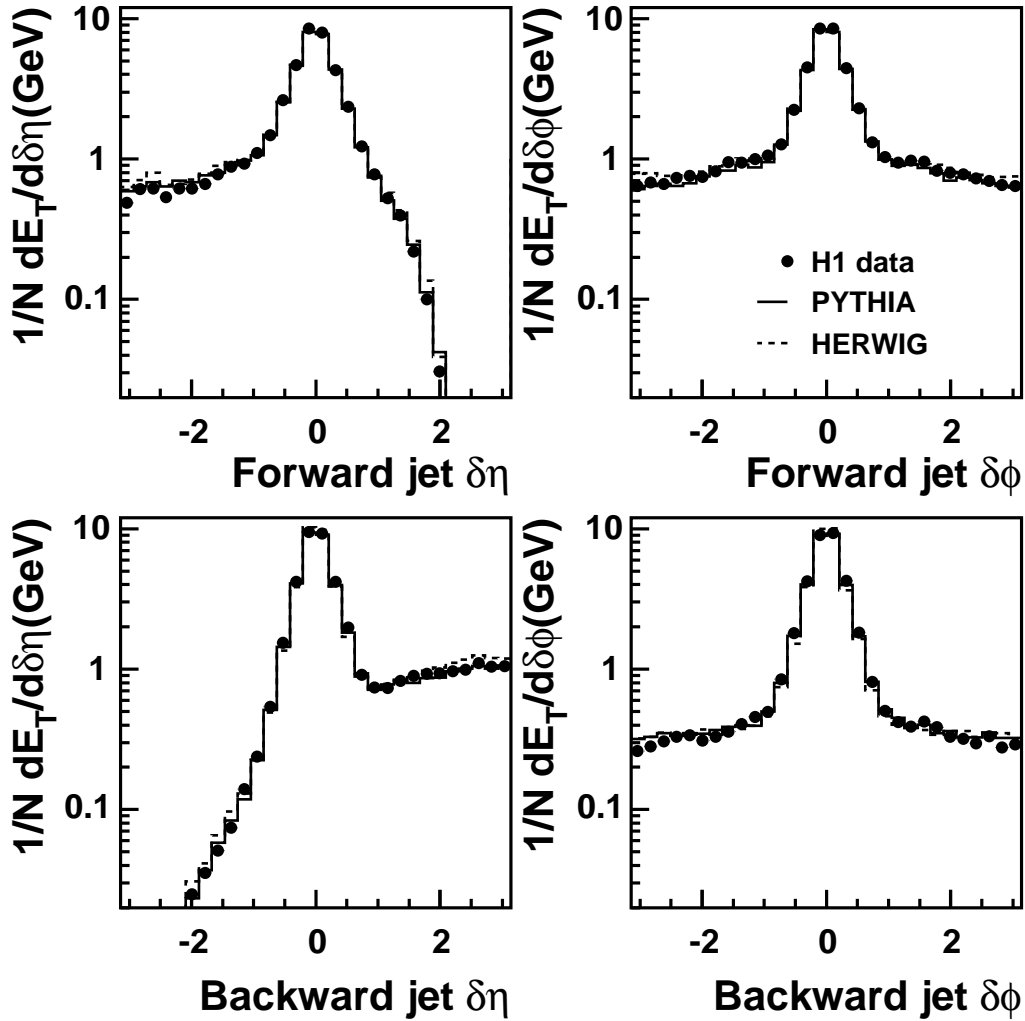,width=\textwidth}
\end{center}
\caption{Jet profiles represented by the energy flow per jet per unit 
rapidity ($\delta \eta$) and per unit azimuth ($\delta \phi$) around 
the jet axis for jets forward and backward relative to the proton 
direction. The H1 data, uncorrected for detector effects, are shown as points. The solid histogram is the prediction of the \pythia~simulation, and the dashed histogram that of \herwig, after being passed through a full simulation of the H1 detector.\label{profiles}}
\end{figure}

There is a large uncertainty in the knowledge of the gap formation probability in standard photoproduction processes; \pythia~produces fewer gaps than \herwig~+ the multiple interactions package \jimmy. This difference is primarily due to the different hadronisation models employed in the two generators. The predictions of \pythia, which exhibit an exponential fall of the gap fraction as $\Delta \eta$ increases, are more in line with naive expectations, although there is no a priori reason to discount the cluster fragmentation model of \herwig. Finally, even within a particular model, the gap formation probability is of course dependent upon the treatment of multiple interactions, although the requirement that the level of hadronic activity in the event matches the data, as shown in figure \ref{profiles}, provides some contraints.

\section{Summary}
\label{summarise}

Dijet events in photoproduction have been studied in which there is a large rapidity separation between the two highest $E_T$ jets. The inclusive dijet cross sections have been measured as functions of the longitudinal momentum fractions of the photon and proton which participate in the production of the jets, $\xgjet$ and $x_p^{jets}$ respectively, $\Delta \eta$, the pseudorapidity separation between the two highest $E_T$ jets, and $\etgap$, the total summed transverse energy between the jets. A significant excess of dijet events with small $\etgap$ is observed for $\Delta \eta > 2.5$ over that predicted by standard photoproduction models. 

In order to investigate the excess, the dijet cross sections have been measured with the additional constraint that the transverse energy between the jets be less than $0.5$, $1.0$, $1.5$ and $2.0$ GeV. The ratios of these cross sections to the inclusive dijet cross sections, the gap fractions, are found to be reasonably well described with the addition of a colour singlet exchange component in the form of the LLA BFKL pomeron, although there is little sensitivity in the data to the underlying dynamics of the model.      

%%%%%%%%%%%%%%%%%%%%%%%%%%%%%%%%%%%%%%%%%%%%%%%%%%%%%%%%%%%%
\section*{Acknowledgements}

We thank Jeff Forshaw and Mike Seymour for many useful discussions and suggestions. We are grateful to the HERA machine group whose outstanding
efforts have made and continue to make this experiment possible. 
We thank
the engineers and technicians for their work in constructing and now
maintaining the H1 detector, our funding agencies for 
financial support, the
DESY technical staff for continual assistance, 
and the DESY directorate for the
hospitality which they extend to the non DESY 
members of the collaboration.

%%%%%%%%%%%%%%%%%%%%%%%%%%%%%%%%%%%%%%%%%%%%%%%%%%%%%%%%%%%%

\newpage
\begin{table}
\begin{center}
\begin{tabular}{|c||c|c|c|c|} \hline
 $\xgjet$ & ${\rm d}\sigma/{\rm d}\xgjet$ & $\delta_{stat}$ & $\delta_{uncor}$ & $\delta_{corr}$  \\
 & (nb) & (nb) & (nb) & (nb) \\  \hline
  0.30 - 0.60  &  1.29 & 0.03  &  0.19 & 0.28   \\ \hline   
  0.60 - 0.75  &  2.27 & 0.06  &  0.16 & 0.41   \\ \hline   
  0.75 - 0.90  &  2.53 & 0.07  &  0.29 & 0.31   \\ \hline   
  0.90 - 1.00  &  0.68 & 0.05  &  0.07 & 0.07   \\ \hline   
 \hline   
 $\xpjet$ & ${\rm d}\sigma/{\rm d}\xpjet$ & $\delta_{stat}$ & $\delta_{uncor}$ & $\delta_{corr}$  \\
 & (nb) & (nb) & (nb) & (nb) \\  \hline
  0.02 - 0.04  &  22.3 & 0.5  &  2.0 & 3.9   \\ \hline   
  0.04 - 0.06  &  23.2 & 0.6  &  2.2 & 3.9   \\ \hline   
  0.06 - 0.08  &  08.6 & 0.3  &  0.9 & 1.7   \\ \hline   
  0.08 - 0.10  &  02.8 & 0.2  &  0.5 & 0.8   \\ \hline   
 \hline   
 $\Delta \eta$ & ${\rm d}\sigma/{\rm d}\Delta \eta$ & $\delta_{stat}$ & $\delta_{uncor}$ & $\delta_{corr}$  \\
 & (nb) & (nb) & (nb) & (nb) \\  \hline
  2.5 - 2.8  &  1.72 & 0.04  &  0.14 & 0.28   \\ \hline   
  2.8 - 3.1  &  1.16 & 0.03  &  0.09 & 0.21   \\ \hline   
  3.1 - 3.5  &  0.67 & 0.02  &  0.08 & 0.13   \\ \hline   
  3.5 - 4.0  &  0.17 & 0.01  &  0.02 & 0.03   \\ \hline   
 \hline   
 $\etgap$ & ${\rm d}\sigma/{\rm d}\etgap$ & $\delta_{stat}$ & $\delta_{uncor}$ & $\delta_{corr}$  \\
 (GeV) & (nb/GeV) & (nb/GeV) & (nb/GeV) & (nb/GeV) \\  \hline
  0.0 - 0.50  &  0.122 & 0.013  &  0.016 & 0.015   \\ \hline   
  0.5 - 1.50  &  0.089 & 0.006  &  0.013 & 0.008   \\ \hline   
  1.5 - 3.50  &  0.141 & 0.005  &  0.027 & 0.016   \\ \hline   
  3.5 - 7.00  &  0.124 & 0.003  &  0.014 & 0.018   \\ \hline   
  7.0 - 12.0  &  0.054 & 0.001  &  0.009 & 0.012   \\ \hline   
\end{tabular}
\end{center}
\caption{The inclusive dijet cross sections differential in $\xgjet$, $x_p^{jets}$, $\Delta \eta$ and $\etgap$. Also shown are the statistical error, $\delta_{stat}$, uncorrelated systematic error,
$\delta_{uncor}$, and correlated systematic error, $\delta_{corr}$.}
\label{inctable}
\end{table}

\begin{table}
\begin{center}
\begin{tabular}{|c|c||c|c|c|c||c|c|c|} \hline
$\etcut$ & $\Delta \eta$ & ${\rm d}\sigma/{\rm d}\Delta \eta$ & $\delta_{stat}$ & $\delta_{uncor}$ & $\delta_{corr}$ & $f(\Delta \eta)$  &  $\delta_{stat}$ & $\delta_{syst}$ \\
 (GeV) & &  (nb) & (nb) & (nb) & (nb) & & &        \\ \hline
 0.5  &  2.5 - 2.8  &  0.085 & 0.011 &  0.023 & 0.012  & 0.050 &  0.006 & 0.018    \\ \hline    
 0.5  &  2.8 - 3.1  &  0.065 & 0.014 &  0.024 & 0.009  & 0.056 &  0.011 & 0.019    \\ \hline    
 0.5  &  3.1 - 3.5  &  0.017 & 0.006 &  0.003 & 0.002  & 0.025 &  0.009 & 0.010    \\ \hline    
 0.5  &  3.5 - 4.0  &  0.017 & 0.007 &  0.004 & 0.002  & 0.100 &  0.034 & 0.039    \\ \hline    
 1.0  &  2.5 - 2.8  &  0.146 & 0.014 &  0.029 & 0.019  & 0.085 &  0.008 & 0.023    \\ \hline    
 1.0  &  2.8 - 3.1  &  0.091 & 0.014 &  0.030 & 0.011  & 0.079 &  0.011 & 0.022    \\ \hline    
 1.0  &  3.1 - 3.5  &  0.046 & 0.010 &  0.004 & 0.005  & 0.069 &  0.014 & 0.019    \\ \hline    
 1.0  &  3.5 - 4.0  &  0.017 & 0.006 &  0.004 & 0.002  & 0.101 &  0.029 & 0.035    \\ \hline    
 1.5  &  2.5 - 2.8  &  0.234 & 0.019 &  0.030 & 0.021  & 0.137 &  0.010 & 0.027    \\ \hline    
 1.5  &  2.8 - 3.1  &  0.157 & 0.018 &  0.058 & 0.020  & 0.136 &  0.014 & 0.038    \\ \hline    
 1.5  &  3.1 - 3.5  &  0.066 & 0.011 &  0.006 & 0.007  & 0.099 &  0.014 & 0.024    \\ \hline    
 1.5  &  3.5 - 4.0  &  0.020 & 0.005 &  0.002 & 0.002  & 0.114 &  0.027 & 0.032    \\ \hline    
 2.0  &  2.5 - 2.8  &  0.305 & 0.020 &  0.038 & 0.030  & 0.178 &  0.010 & 0.033    \\ \hline    
 2.0  &  2.8 - 3.1  &  0.193 & 0.019 &  0.054 & 0.022  & 0.167 &  0.014 & 0.039    \\ \hline    
 2.0  &  3.1 - 3.5  &  0.092 & 0.012 &  0.013 & 0.010  & 0.137 &  0.016 & 0.027    \\ \hline    
 2.0  &  3.5 - 4.0  &  0.033 & 0.007 &  0.006 & 0.005  & 0.189 &  0.034 & 0.044    \\ \hline    
\end{tabular}
\end{center}
\caption{The dijet cross sections differential in $\Delta \eta$, with the additional requirement that $\etgap < \etcut$, shown with the statistical error, $\delta_{stat}$, uncorrelated systematic error,
$\delta_{uncor}$, and correlated systematic error, $\delta_{corr}$. Also shown are the gap fractions, $f(\Delta \eta)$, defined as the fraction of all dijet events with $\etgap < \etcut$, and their
associated statistical and systematic errors.}
\label{deteatable}
\end{table}

\begin{table}
\begin{center}
\begin{tabular}{|c|c||c|c|c|c||c|c|c|} \hline
$\etcut$ & $\xgjet$ & ${\rm d}\sigma/{\rm d}\xgjet$ & $\delta_{stat}$ & $\delta_{uncor}$ & $\delta_{corr}$ & $f(\xgjet)$  &  $\delta_{stat}$ & $\delta_{syst}$ \\
 (GeV) & &  (nb) & (nb) & (nb) & (nb) & & &        \\ \hline
 0.5  &  0.30 - 0.60  &  0.031 & 0.009 &  0.005 & 0.004  & 0.024 &  0.007 & 0.009    \\ \hline    
 0.5  &  0.60 - 0.75  &  0.105 & 0.027 &  0.058 & 0.017  & 0.046 &  0.011 & 0.021    \\ \hline    
 0.5  &  0.75 - 0.90  &  0.115 & 0.022 &  0.022 & 0.014  & 0.045 &  0.008 & 0.013    \\ \hline    
 0.5  &  0.90 - 1.00  &  0.188 & 0.030 &  0.040 & 0.027  & 0.277 &  0.035 & 0.065    \\ \hline    
 1.0  &  0.30 - 0.60  &  0.044 & 0.009 &  0.006 & 0.006  & 0.034 &  0.007 & 0.010    \\ \hline    
 1.0  &  0.60 - 0.75  &  0.135 & 0.026 &  0.049 & 0.018  & 0.060 &  0.011 & 0.020    \\ \hline    
 1.0  &  0.75 - 0.90  &  0.195 & 0.026 &  0.030 & 0.020  & 0.077 &  0.010 & 0.017    \\ \hline    
 1.0  &  0.90 - 1.00  &  0.313 & 0.037 &  0.054 & 0.046  & 0.462 &  0.034 & 0.068    \\ \hline    
 1.5  &  0.30 - 0.60  &  0.080 & 0.013 &  0.027 & 0.010  & 0.062 &  0.009 & 0.019    \\ \hline    
 1.5  &  0.60 - 0.75  &  0.178 & 0.027 &  0.061 & 0.019  & 0.079 &  0.011 & 0.024    \\ \hline    
 1.5  &  0.75 - 0.90  &  0.373 & 0.037 &  0.052 & 0.029  & 0.147 &  0.013 & 0.025    \\ \hline    
 1.5  &  0.90 - 1.00  &  0.423 & 0.041 &  0.056 & 0.054  & 0.623 &  0.028 & 0.065    \\ \hline    
 2.0  &  0.30 - 0.60  &  0.111 & 0.015 &  0.047 & 0.017  & 0.086 &  0.011 & 0.028    \\ \hline    
 2.0  &  0.60 - 0.75  &  0.223 & 0.028 &  0.054 & 0.033  & 0.098 &  0.012 & 0.023    \\ \hline    
 2.0  &  0.75 - 0.90  &  0.520 & 0.040 &  0.069 & 0.044  & 0.205 &  0.013 & 0.031    \\ \hline    
 2.0  &  0.90 - 1.00  &  0.483 & 0.042 &  0.072 & 0.060  & 0.712 &  0.024 & 0.078    \\ \hline    
\end{tabular}
\end{center}
\caption{The dijet cross sections differential in $\xgjet$, with the additional requirement that $\etgap < \etcut$, shown with the statistical error, $\delta_{stat}$, uncorrelated systematic error,
$\delta_{uncor}$, and correlated systematic error, $\delta_{corr}$. Also shown are the gap fractions,  $f(\xgjet)$, defined as
the fraction of all dijet events with $\etgap < \etcut$, and their
associated statistical and systematic errors.}
\label{xgtable}
\end{table}

\begin{table}
\begin{center}
\begin{tabular}{|c|c||c|c|c|c||c|c|c|} \hline
$\etcut$ & $\xpjet$ & ${\rm d}\sigma/{\rm d}\xpjet$ & $\delta_{stat}$ & $\delta_{uncor}$ & $\delta_{corr}$ & $f(\xpjet)$  &  $\delta_{stat}$ & $\delta_{syst}$ \\
 (GeV) & &  (nb) & (nb) & (nb) & (nb) & & &        \\ \hline
 0.5  &  0.02 - 0.04  &  1.02 & 0.17 &  0.22 & 0.09  & 0.046 &  0.007 & 0.013    \\ \hline    
 0.5  &  0.04 - 0.06  &  1.01 & 0.21 &  0.18 & 0.12  & 0.044 &  0.009 & 0.014    \\ \hline    
 0.5  &  0.06 - 0.08  &  0.34 & 0.14 &  0.06 & 0.04  & 0.040 &  0.016 & 0.018    \\ \hline    
 0.5  &  0.08 - 0.10  &  0.10 & 0.07 &  0.03 & 0.01  & 0.034 &  0.024 & 0.028    \\ \hline    
 1.0  &  0.02 - 0.04  &  2.27 & 0.24 &  0.48 & 0.23  & 0.102 &  0.010 & 0.021    \\ \hline    
 1.0  &  0.04 - 0.06  &  1.21 & 0.20 &  0.32 & 0.14  & 0.052 &  0.008 & 0.023    \\ \hline    
 1.0  &  0.06 - 0.08  &  0.46 & 0.14 &  0.07 & 0.05  & 0.054 &  0.015 & 0.019    \\ \hline    
 1.0  &  0.08 - 0.10  &  0.16 & 0.07 &  0.04 & 0.02  & 0.057 &  0.024 & 0.030    \\ \hline    
 1.5  &  0.02 - 0.04  &  3.67 & 0.32 &  1.06 & 0.43  & 0.165 &  0.012 & 0.036    \\ \hline    
 1.5  &  0.04 - 0.06  &  1.92 & 0.25 &  0.24 & 0.15  & 0.083 &  0.010 & 0.020    \\ \hline    
 1.5  &  0.06 - 0.08  &  0.54 & 0.12 &  0.10 & 0.09  & 0.063 &  0.014 & 0.020    \\ \hline    
 1.5  &  0.08 - 0.10  &  0.16 & 0.07 &  0.03 & 0.01  & 0.057 &  0.024 & 0.034    \\ \hline    
 2.0  &  0.02 - 0.04  &  4.52 & 0.33 &  0.79 & 0.48  & 0.202 &  0.012 & 0.033    \\ \hline    
 2.0  &  0.04 - 0.06  &  2.91 & 0.30 &  0.40 & 0.32  & 0.125 &  0.011 & 0.021    \\ \hline    
 2.0  &  0.06 - 0.08  &  1.02 & 0.20 &  0.26 & 0.14  & 0.118 &  0.020 & 0.032    \\ \hline    
 2.0  &  0.08 - 0.10  &  0.28 & 0.10 &  0.06 & 0.02  & 0.101 &  0.033 & 0.052    \\ \hline    
\end{tabular}
\end{center}
\caption{The dijet cross sections differential in $\xpjet$, with the additional requirement that $\etgap < \etcut$, shown with the statistical error, $\delta_{stat}$, uncorrelated systematic error,
$\delta_{uncor}$, and correlated systematic error, $\delta_{corr}$. Also shown are the gap fractions,  $f(\xpjet)$, defined as
the fraction of all dijet events with $\etgap < \etcut$, and their
associated statistical and systematic errors.\label{xptable}}
\end{table}

\end{document}

%% file: h1auts.tex
%-- H1AUTS Author list by names 
%-- Status: Thu Aug  2 18:19:24 MET DST 2001  Number of authors = 332 

C.~Adloff$^{33}$,              %WUPP-ST        01/96           Adloff              
V.~Andreev$^{24}$,             %LPI -PD        8/88            Andreev             
B.~Andrieu$^{27}$,             %ECPL-PD        8/88            Andrieu             
T.~Anthonis$^{4}$,             %ANTW-ST        11/99           Anthonis            
V.~Arkadov$^{35}$,             %ZEUT-LEFT      10/0            Arkadov             
A.~Astvatsatourov$^{35}$,      %ZEUT-ST        02/98           Astvatsatourov      
A.~Babaev$^{23}$,              %ITEP-PD        8/88            Babaev              
J.~B\"ahr$^{35}$,              %ZEUT-PD        8/88            Baehr               
P.~Baranov$^{24}$,             %LPI -PD        8/88            Baranovp            
E.~Barrelet$^{28}$
% ...

%-- H1 Institutes 
\bigskip{\it
% ...
$ ^{4}$ Inter-University Institute for High Energies ULB-VUB, Brussels;
          Universiteit Antwerpen (UIA), Antwerpen; Belgium$^{ c}$ \\
% ...
$ ^{23}$ Institute for Theoretical and Experimental Physics,
          Moscow, Russia$^{ l}$ \\
 $ ^{24}$ Lebedev Physical Institute, Moscow, Russia$^{ e}$ \\
% ...
$ ^{27}$ LPNHE, Ecole Polytechnique, IN2P3-CNRS, Palaiseau, France \\
 $ ^{28}$ LPNHE, Universit\'{e}s Paris VI and VII, IN2P3-CNRS,
          Paris, France \\
% ...
$ ^{33}$ Fachbereich Physik, Bergische Universit\"at Gesamthochschule
          Wuppertal, Wuppertal, Germany \\
% ...
$ ^{35}$ DESY, Zeuthen, Germany \\
% ...
$ ^c$ Supported by FNRS-FWO-Vlaanderen, IISN-IIKW and IWT \\
% ...
$ ^e$ Supported by the Deutsche Forschungsgemeinschaft \\
% ...
$ ^l$ Partially Supported by Russian Foundation
      for Basic Research, grant    no. 00-15-96584 \\
}

%% file: paper.bbl
\begin{thebibliography}{99}
%\cite{Forshaw:1997dc}
\bibitem{FR}
See for example J.~R.~Forshaw and D.~A.~Ross,
%``Quantum chromodynamics and the pomeron,''
  Cambridge UK Univ. Pr. (1997) 248 p. (Cambridge lecture notes in physics. 9) and references therein.

%\cite{Adloff:1997sc}
\bibitem{H1diff}
C.~Adloff {\it et al.}  [H1 Collaboration],
%``Inclusive measurement of diffractive deep-inelastic e p scattering''
Z.\ Phys.\ C {\bf 76} (1997) 613
[hep-ex/9708016].
%%CITATION = HEP-EX 9708016;%%

%\cite{Breitweg:1999gc}
\bibitem{ZEUSdiff}
J.~Breitweg {\it et al.}  [ZEUS Collaboration],
%``Measurement of the diffractive cross section in deep inelastic  scattering using ZEUS 1994 data,''
Eur.\ Phys.\ J.\ C {\bf 6} (1999) 43
[hep-ex/9807010].
%%CITATION = HEP-EX 9807010;%%      

%\cite{Abachi:1994hb}
\bibitem{DZero1}
S.~Abachi {\it et al.}  [D0 Collaboration],
%``Rapidity gaps between jets in p anti-p collisions at s**(1/2) = 1.8-TeV,''
Phys.\ Rev.\ Lett.\  {\bf 72} (1994) 2332.
%%CITATION = PRLTA,72,2332;%%                 

%\cite{Abachi:1996gz}
\bibitem{DZero2}
S.~Abachi {\it et al.}  [D0 Collaboration],
%``Jet Production via Strongly-Interacting Color-Singlet Exchange in $p\bar{p}$ Collisions,''
Phys.\ Rev.\ Lett.\  {\bf 76} (1996) 734
[hep-ex/9509013].
%%CITATION = HEP-EX 9509013;%%

%\cite{Abbott:1998jb}
\bibitem{DZero3}
B.~Abbott {\it et al.}  [D0 Collaboration],
%``Probing hard color-singlet exchange in p anti-p collisions at s**(1/2)  = 630-GeV and 1800-GeV,''
Phys.\ Lett.\ B {\bf 440} (1998) 189
[hep-ex/9809016].
%%CITATION = HEP-EX 9809016;%%

%\cite{Abe:1995de}
\bibitem{CDF1}
F.~Abe {\it et al.}  [CDF Collaboration],
%``Observation of rapidity gaps in anti-p p collisions at 1.8-TeV,''
Phys.\ Rev.\ Lett.\  {\bf 74} (1995) 855.
%%CITATION = PRLTA,74,855;%%

%\cite{Abe:1998ie}
\bibitem{CDF2}
F.~Abe {\it et al.}  [CDF Collaboration],
%``Dijet production by color-singlet exchange at the Fermilab Tevatron,''
Phys.\ Rev.\ Lett.\  {\bf 80} (1998) 1156.
%%CITATION = PRLTA,80,1156;%%

%\cite{Abe:1998ip}
\bibitem{CDF3}
F.~Abe {\it et al.}  [CDF Collaboration],
%``Events with a rapidity gap between jets in anti-p p collisions at  s**(1/2) = 630-GeV,''
Phys.\ Rev.\ Lett.\  {\bf 81} (1998) 5278.
%%CITATION = PRLTA,81,5278;%%

%\cite{Derrick:1996pb}
\bibitem{zeus}
M.~Derrick {\it et al.}  [ZEUS Collaboration],
%``Rapidity Gaps between Jets in Photoproduction at HERA,''
Phys.\ Lett.\ B {\bf 369} (1996) 55
[hep-ex/9510012].
%%CITATION = HEP-EX 9510012;%%

%\cite{Forshaw:1998wn}
\bibitem{FS}
J.~R.~Forshaw and P.~J.~Sutton,
%``Diffusion and the BFKL pomeron,''
Eur.\ Phys.\ J.\ C {\bf 1} (1998) 285
[hep-ph/9703225].
%%CITATION = HEP-PH 9703225;%%

%\cite{Kuraev:1977fs}
\bibitem{BFKL1}
E.~A.~Kuraev, L.~N.~Lipatov and V.~S.~Fadin,
%``The Pomeranchuk Singularity In Nonabelian Gauge Theories,''
Sov.\ Phys.\ JETP {\bf 45} (1977) 199
[Zh.\ Eksp.\ Teor.\ Fiz.\  {\bf 72} (1977) 377].
%%CITATION = SPHJA,45,199;%%

%\cite{Balitsky:1978ic}
\bibitem{BFKL2}
I.~I.~Balitsky and L.~N.~Lipatov,
%``The Pomeranchuk Singularity In Quantum Chromodynamics,''
Sov.\ J.\ Nucl.\ Phys.\  {\bf 28} (1978) 822.
[Yad.\ Fiz.\  {\bf 28} (1978) 1597].
%%CITATION = SJNCA,28,822;%%

%\cite{Lipatov:1986uk}
\bibitem{BFKL3}
L.~N.~Lipatov,
%``The Bare Pomeron In Quantum Chromodynamics,''
Sov.\ Phys.\ JETP {\bf 63} (1986) 904.
[Zh.\ Eksp.\ Teor.\ Fiz.\  {\bf 90} (1986) 1536].
%%CITATION = SPHJA,63,904;%%

%\cite{Mueller:1992pe}
\bibitem{MT}
A.~H.~Mueller and W.~K.~Tang,
%``High-energy parton-parton elastic scattering in QCD,''
Phys.\ Lett.\ B {\bf 284} (1992) 123.
%%CITATION = PHLTA,B284,123;%%

%\cite{Enberg:2001ev}
\bibitem{enberg}
R.~Enberg, G.~Ingelman and L.~Motyka,
%``Hard colour singlet exchange and gaps between jets at the Tevatron,''
Phys.\ Lett.\ B {\bf 524} (2002) 273
[hep-ph/0111090].
%%CITATION = HEP-PH 0111090;%%

%\cite{Cox:1999dw}
\bibitem{CFL}
B.~Cox, J.~Forshaw and L.~L\"onnblad,
%``Hard colour singlet exchange at the Tevatron,''
JHEP {\bf 9910} (1999) 023
[hep-ph/9908464].
%%CITATION = HEP-PH 9908464;%%

%\cite{Bjorken:1993er}
\bibitem{Bj}
J.~D.~Bjorken,
%``Rapidity gaps and jets as a new physics signature in very high-energy hadron-hadron collisions,''
Phys.\ Rev.\ D {\bf 47} (1993) 101.
%%CITATION = PHRVA,D47,101;%%

\bibitem{fsel}
R.S.~Fletcher and T.~Stelzer, Phys. Rev. {\bf D48} (1993) 5162.

%\cite{Gotsman:1998mm}
\bibitem{GLM}
E.~Gotsman, E.~Levin and U.~Maor,
%``Energy dependence of the survival probability of large rapidity gaps,''
Phys.\ Lett.\ B {\bf 438} (1998) 229
[hep-ph/9804404].
%%CITATION = HEP-PH 9804404;%%

%\cite{Kaidalov:2001iz}
\bibitem{Kaidalov:2001iz}
A.~B.~Kaidalov, V.~A.~Khoze, A.~D.~Martin and M.~G.~Ryskin,
%``Probabilities of rapidity gaps in high energy interactions,''
Eur.\ Phys.\ J.\ C {\bf 21} (2001) 521
[hep-ph/0105145].
%%CITATION = HEP-PH 0105145;%%

%\cite{Oderda:1999ta}
\bibitem{OS}
G.~Oderda and G.~Sterman,
%``Interjet rapidity gaps in perturbative QCD,''
[hep-ph/9910414].
%%CITATION = HEP-PH 9910414;%%

%\cite{Oderda:2000kr}
\bibitem{OShera}
G.~Oderda,
%``Dijet rapidity gaps in photoproduction from perturbative {QCD},''
Phys.\ Rev.\ D {\bf 61} (2000) 014004
[hep-ph/9903240].
%%CITATION = HEP-PH 9903240;%%

%\cite{Eboli:1999dd}
\bibitem{Eboli:1999dd}
O.~J.~Eboli, E.~M.~Gregores and F.~Halzen,
%``Color evaporation induced rapidity gaps,''
Phys.\ Rev.\ D {\bf 61} (2000) 034003
[hep-ph/9908374].
%%CITATION = HEP-PH 9908374;%%

%\cite{Abt:1997hi}
\bibitem{h1nim}
I.~Abt {\it et al.}  [H1 Collaboration],
%``The H1 detector at HERA,''
Nucl.\ Instrum.\ Meth.\ A {\bf 386} (1997) 310 and 348.

\bibitem{spac}
H1 SPACAL group, R.D.~Appuhn et al., Nucl. Instr. and Meth. {\bf A386}
(1997) 397.
%%CITATION = NUIMA,A386,310;%%

\bibitem{fscomb}
C.~Adloff {\it et al.}  [H1 Collaboration],
Z.\ Phys. \ C {\bf 74} (1997) 221
[hep-ph/9702003].
%%CITATION = HEP-EX 9702003;%%

%\cite{Ellis:tq}
\bibitem{Ellis:tq}
S.~D.~Ellis and D.~E.~Soper,
%``Successive Combination Jet Algorithm For Hadron Collisions,''
Phys.\ Rev.\ D {\bf 48} (1993) 3160
[hep-ph/9305266].
%%CITATION = HEP-PH 9305266;%%

%\cite{Catani:1993hr}
\bibitem{kt}
S.~Catani, Y.~L.~Dokshitzer, M.~H.~Seymour and B.~R.~Webber,
%``Longitudinally invariant K(t) clustering algorithms for hadron-hadron collisions,''
Nucl.\ Phys.\ B {\bf 406} (1993) 187.
%%CITATION = NUPHA,B406,187;%%

\bibitem{frixione}
S.~Frixione and G.~Ridolfi,
%``Jet photoproduction at HERA,''
Nucl.\ Phys.\ B {\bf 507} (1997) 315
[hep-ph/9707345].
%%CITATION = HEP-PH 9707345;%%

\bibitem{PYTHIA} T.~Sj\"ostrand, %``High-energy physics event generation
with PYTHIA 5.7 and JETSET 7.4, Comput.\ Phys.\ Commun.\ {\bf 82} (1994)
74.
%%CITATION = CPHCB,82,74;%%

\bibitem{HERWIG}
G.~Marchesini, B.~R.~Webber, G.~Abbiendi, I.~G.~Knowles, M.~H.~Seymour and L.~Stanco,
%``HERWIG: A Monte Carlo event generator for simulating hadron emission reactions with interfering gluons. Version 5.1 - April 1991,''
Comput.\ Phys.\ Commun.\  {\bf 67} (1992) 465.
%%CITATION = CPHCB,67,465;%%

\bibitem{jimmy}
J.~M.~Butterworth, J.~R.~Forshaw and M.~H.~Seymour,
%``Multiparton interactions in photoproduction at HERA,''
Z.\ Phys.\ C {\bf 72} (1996) 637
[hep-ph/9601371].
%%CITATION = HEP-PH 9601371;%%

\bibitem{GRV}
M.~Gl\"uck, E.~Reya and A.~Vogt,
%``Dynamical parton distributions of the proton and small x physics,''
Z.\ Phys.\ C {\bf 67} (1995) 433.
%%CITATION = ZEPYA,C67,433;%%

%\cite{Gluck:1991jc}
\bibitem{Gluck:1991jc}
M.~Gl\"uck, E.~Reya and A.~Vogt,
%``Photonic parton distributions,''
Phys.\ Rev.\ D {\bf 46} (1992) 1973.
%%CITATION = PHRVA,D46,1973;%%





\end{thebibliography}
